\documentclass{llncs}
\usepackage{epsfig,pifont,xspace,wrapfig,subfigure}
\newcommand{\EA}[1]{{\it et al.}~\cite{#1}}
\newcommand{\FI}[1]{Fig.~\ref{#1}\xspace}
\newcommand{\TA}[1]{Table~\ref{#1}\xspace}
\newcommand{\Sac}{{\em Saccharomyces cerevisiae}\xspace}
\newcommand{\lscc}{{\sc Lscc}\xspace}
\newcommand{\scc}{\mbox{\sc scc}\xspace}

\title{On Cycles in the Transcription Network of \Sac
}
\titlerunning{Cycles in the Yeast Transcription Network}

\author{Jieun Jeong\inst{1}
}

\institute{Department of Computer Science and Engineering, The Pennsylvania
  State University, University Park, PA 16802, USA,
\email{\tt jijeong@cse.psu.edu}
}
\begin{document}

\maketitle

\begin{abstract}
We investigate the cycles in the transcription network of \Sac.  Unlike a
similar network of {\em Escherichia coli}, it contains many cycles.
We characterize properties of these cycles and their place in the
regulatory mechanism of the cell.

Almost all cycles in the transcription network of \Sac are contained in a
single {\em strongly connected component}, which we call \lscc (L for
``largest''), except for a single cycle of two transcription factors.  
    
Among different physiological conditions, cell cycle has the most significant
relationship with \lscc, as the set of 64 transcription interactions that are
active in all phases of the cell cycle has overlap of 27 with the interactions
of \lscc (of which there are 49).

Conversely, if we remove the interactions that are active in all phases of the
cell cycle (fewer than 1\% of the total), the \lscc would have only three
nodes and 5 edges, 4 of which are active only
in the stress response subnetwork.

\lscc has a special place in the topology of the network
and it can be used to define a natural hierarchy in the network;
in every physiological subnetwork \lscc plays a pivotal role.

Apart from those well-defined conditions, the transcription network of \Sac is
devoid of cycles.  It was observed that two conditions that were studied and
that have no cycles of their own are {\em exogenous}: diauxic shift and DNA
repair, while cell cycle, sporulation are {\em endogenous}.
Perhaps, during the slow recovery phase, the stress response is
{\em endogenous} as well.
\end{abstract}

\section{Background}

Cycles have a central role in control of continuing processes (for an
example, see Hartwell
\cite{H02}).
Therefore we expect the regulatory mechanism of a cell to have many cycles
of interactions.  Only some of these interactions have the form of a
transcription factor (TF for short) regulating expression of a target gene.
Our question is: given that there are
cycles of transcription interactions, are they important in the
regulation of life processes?

Graph properties of the regulatory networks have been reported
in a number of papers.
Shen-Orr \EA{A02}
analyzed the
regulatory networks statically and observed certain characteristic
{\em motifs} that are more frequent than in the random model
and which have functional significance (while other small subgraphs
are significantly less frequent).
Luscombe \EA{L04}
studied the dynamics of the regulatory network of \Sac as it
changes for multiple conditions and proposed 
a method for the statistical analysis of network dynamics. They have found
large changes in the topology of the network and compared it with random
graphs.

We have found that the transcription network of \Sac contains a single
large strongly connected component (a union of overlapping cycles), which
we call \lscc, and
that the topology changes discussed by
Luscombe \EA{L04}
are well reflected
within \lscc, in spite of its small size.

Yu and Gerstein \cite{Y06}
have examined at the structure of regulatory
networks and showed that it exhibits a certain natural hierarchy.
We propose another hierarchical partition of the network: above the \lscc,
the \lscc, below the \lscc and ``parallel'' to the \lscc (see \TA{hier})
and we show that this partition is in some sense natural.

Comparisons of biological networks with random graphs were subject
of methodological investigations of
Barabasi and Albert \cite{B99}
who proposed
a {\em scale-free} model.  This model is rather difficult to apply in
the case of directed graphs that have large asymmetry between edge beginnings
and ends; one can have a separate model for the out-degrees --- a power law,
and for the in-degrees --- a Poisson distribution, but parameters of
such graphs converge very slowly, so a model based on such parameters
can be misleading.
Therefore
Milo \EA{M03}
(see also Newman \EA{N01}) proposed several methods of
generating graphs that have the same in- and out-degrees of the reference
network.  We used the faster and somewhat biased variant of their
``matching algorithm''.

\section{Results and Discussion}

In the data set of
Luscombe \EA{L04}
we can see the \lscc
with 25 TFs and one small strongly connected component with two TFs.

To see if the cycles of the \lscc are significant, we checked how the
topological
changes of the transcription network during various physiological conditions
are reflected inside the \lscc, we checked several graph characteristics
of the TFs in the \lscc, and we compared the characteristics of the \lscc
to the cycles in random networks.  

\subsection{General characterization of the cycles}

\subsubsection{Size of \lscc in the expected range.}
The cycles form two connected components, one ``small'',
consisting of 2 TFs, and one ``large'', consisting of 25 TFs.

The degenerate component consists of two TFs with indistinguishable
interactions that have self-loops, thus they are TFs of
themselves, and of each other.  This may be a result of a relatively
recent gene duplication.  Thus we will ignore this cycle in our discussions.

The size of the cyclic component is within the range of variability for
random models, and this range, 17-40 (with the average of 30)
does not change much when we boost
the number of elementary motifs.  Thus the size alone would allow the
cyclic component to be a random artifact of other properties of the network.

It is also typical that there are very few cycles outside the largest
strongly connected component: the average sum of sizes of other
non-trivial strongly connected components is 1.4.

By the way of contrast, the transcription network of {\em Escherichia coli} is
either devoid of cycles or it contains very few of them (depending on the
data set, see Cosentino Lagomarsino \EA{La07}).
\subsubsection{\lscc connected very strongly to the cell cycle.}
The transcription network reported by
Luscombe \EA{L04}
has 142 TFs
and 7074 interactions, of which we disregard 21 ``self-loop'' interactions.
25 TFs and 49 interactions form the \lscc (as they cannot be contained in
longer simple cycles).
The subnetworks associated with the 5 stages of the cell cycle have
64 interactions in common (we name this set {\sc Ccc}, ``common to
cell cycle''), of which 27 are present in the \lscc.
If even one of these two sets of interactions were random, the expected
number of common elements would be smaller than 1 ($49\times 64 / 7053$).

Another way to illustrate how strongly the transcription cycles are
connected to the cell cycle is to define the following sets of
interactions:
{\sc Ai}, all interactions, 7053 elements,
{\sc Ccc}, 64 elements,
{\sc Pccc} (for proper {\sc Ccc} ), {\sc Ccc} without interactions
common to all conditions, 50 elements.

The number of TFs in cycles of interactions
for the set {\sc Ai} is 27,
which is close to the average value of 31.4 obtained in random
tests.  Because {\sc Ccc} and {\sc Pccc} are so small, the tests
for
{\sc Ai}$-${\sc Pccc} and
{\sc Ai}$-${\sc Ccc} 
should have very similar average values, but the actual number
drops from 27 to 8 and 5 respectively.

\subsubsection{Cycles of subnetworks other than cell cycle.}
\newcommand{\Lsc}[1]{\mbox{\sc Lscc}_{\bf #1}}

For subnetwork $A$ we define \lscc$\!\!_A$ as the set of interactions of $A$
that are also in \lscc; to measure the difference between two sets we
use $|A\oplus B|$, the number of elements that are in one of the sets $A,B$
but not in both.

When we compare the subnetworks of the cell cycle and sporulation, we
observe that
$\Lsc{sp}\subset\Lsc{cc}$ and $|\Lsc{sp}\oplus\Lsc{cc}|=12$.
Nevertheless, the cycles of $\Lsc{sp}$ involve only 7 interactions.

In terms of $|A \oplus B|$, stress response is most distant from the cell
cycle: $|\Lsc{sr}\oplus\Lsc{cc}|=32$, as
$|\Lsc{cc}-\Lsc{sr}|=22$ and $|\Lsc{sr}-\Lsc{cc}|=10$.

Stress reponse is also special in the sense that it has cycles of its own,
all of which involve TF YAP6 that is not active
in any other subnetwork.  It seems that the cyclic interaction of this
TF with two other TFs is a differentiating part of stress response
condition.  Two other conditions, diauxic shift and DNA damage, have similar
sets of active interactions in \lscc, but they lack 5 interactions
involving YAP6.

One cycle consists of 3 interactions that are common to all conditions,
REB1 $\rightarrow$ SIN3 $\rightarrow$ HSF1 $\rightarrow$ REB1.  Note that HSF1
is a Heat Stress Factor, very important in the stress response, but also
in ``basal level sustained transcription'' (see Mager and Ferreira \cite{M93}).
One possible
role of cycles in stress response is slowing down the recovery transition 
from the stress condition, so it can last several hours \cite{M93}.
During the recovery, sporulation and cell cycle activities are supressed.
In this sense, stress response is partially {\em endogenous} to use
the classification of
Luscombe \EA{L04}
(they group Cell Cycle and
Sporulation as endogenous and the other conditions as exogenous).

\subsubsection{\lscc has an orderly layout.}

\begin{figure}[t]
\begin{picture}(340,220)(0,120)
\put(152,110){\epsfig{file=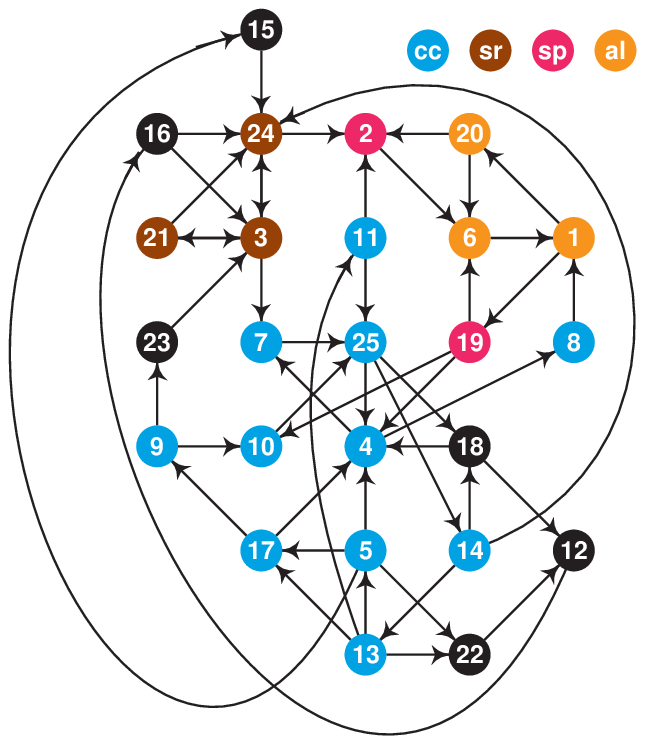,width=2.7in}}

\put(-5,330){\footnotesize node}
\put(25,330){\footnotesize code}
\put(60,330){\footnotesize protein}

\put( 0,315){\footnotesize 1}
\put(10,315){\footnotesize YBR049C}
\put(60,315){\footnotesize REB1}

\put( 0,305){\footnotesize 2}
\put(10,305){\footnotesize YDR207C}
\put(60,305){\footnotesize UME6}

\put( 0,295){\footnotesize 3}
\put(10,295){\footnotesize YDR259C}
\put(60,295){\footnotesize YAP6}

\put( 0,285){\footnotesize 4}
\put(10,285){\footnotesize YDR501W }
\put(60,285){\footnotesize PLM2}

\put( 0,275){\footnotesize 5}
\put(10,275){\footnotesize YER111C }
\put(60,275){\footnotesize SWI4}

\put( 0,265){\footnotesize 6}
\put(10,265){\footnotesize YGL073W }
\put(60,265){\footnotesize HSF1}

\put( 0,255){\footnotesize 7}
\put(10,255){\footnotesize YIL122W }
\put(60,255){\footnotesize POG1}

\put( 0,245){\footnotesize 8}
\put(10,245){\footnotesize YJR060W }
\put(60,245){\footnotesize CBF1}

\put( 0,235){\footnotesize 9}
\put(10,235){\footnotesize YKL043W }
\put(60,235){\footnotesize PHD1}

\put(-5,225){\footnotesize 10}
\put(10,225){\footnotesize YKL062W }
\put(60,225){\footnotesize MSN4}

\put(-5,215){\footnotesize 11}
\put(10,215){\footnotesize YKL112W }
\put(60,215){\footnotesize ABF1}

\put(-5,205){\footnotesize 12}
\put(10,205){\footnotesize YLR131C }
\put(60,205){\footnotesize ACE2}

\put(-5,195){\footnotesize 13}
\put(10,195){\footnotesize YLR182W }
\put(60,195){\footnotesize SWI6}

\put(-5,185){\footnotesize 14}
\put(10,185){\footnotesize YLR183C }
\put(60,185){\footnotesize TOS4}

\put(-5,175){\footnotesize 15}
\put(10,175){\footnotesize YLR256W }
\put(60,175){\footnotesize HAP1}

\put(-5,165){\footnotesize 16}
\put(10,165){\footnotesize YML007W }
\put(60,165){\footnotesize YAP1}

\put(-5,155){\footnotesize 17}
\put(10,155){\footnotesize YML027W }
\put(60,155){\footnotesize YOX1}

\put(-5,145){\footnotesize 18}
\put(10,145){\footnotesize YNL068C }
\put(60,145){\footnotesize FKH2}

\put(-5,135){\footnotesize 19}
\put(10,135){\footnotesize YNL216W }
\put(60,135){\footnotesize RAP1}

\put(-5,125){\footnotesize 20}
\put(10,125){\footnotesize YOL004W }
\put(60,125){\footnotesize SIN3}

\put(-5,115){\footnotesize 21}
\put(10,115){\footnotesize YOR028C }
\put(60,115){\footnotesize CIN5 }

\put(-5,105){\footnotesize 22}
\put(10,105){\footnotesize YOR372C }
\put(60,105){\footnotesize NDD1}

\put(95,125){\footnotesize 23}
\put(110,125){\footnotesize YPL177C }
\put(160,125){\footnotesize CUP9}

\put(95,115){\footnotesize 24}
\put(110,115){\footnotesize YPR065W }
\put(160,115){\footnotesize ROX1}

\put(95,105){\footnotesize 25}
\put(110,105){\footnotesize YPR104C }
\put(160,105){\footnotesize FHL1}
\end{picture}
\vspace{3ex}
\caption{\label{allfigure}The diagram of \lscc}.
\end{figure}
\FI{allfigure} shows the graph formed by the transcription factors and
interactions of \lscc, with
nodes placed on a square grid as to minimize the edge lengths.  Note that
rather few edges (7 of 49) are longer than a single square side/diagonal.

In the diagram, {\bf al} (apricot color) marks the nodes present in the
cycles of all subnetworks.  The cycles in the diauxic shift and DNA damage
subnetworks contain only these nodes.  (Note that an interaction of \lscc
can be active in a subnetwork without belonging to a cycle in that
subnetwork.)

The cycles in the sporulation subnetwork {\bf sp}
contain apricot and strawberry nodes.

The cycles in the cell cycle subnetwork {\bf cc} contain apricot, strawberry
and cerulean nodes.

The cycles in the stress response subnetwork {\bf sr}
contain apricot and sienna nodes.

One can note that the graph does not appear random.
One feature is that it can be laid on a regular grid with few long edges.
The second is that functionally defined groups of nodes, {\bf al, sp, cc, sr}
and {\bf ds} are rather well separated from each other.

\subsubsection{\lscc has small feedback vertex set and three natural subunits.}

\newcommand{\fvs}{2(a)}
\newcommand{\units}{2(b)}
\newcommand{\changeso}{3}
\newcommand{\changest}{4}
\begin{figure}[t]
\centerline{
\subfigure[\label{fvs}A pictorial proof that $\{1,3,25\}$ is the unique minimum feedback
vertex set.]{\epsfig{file=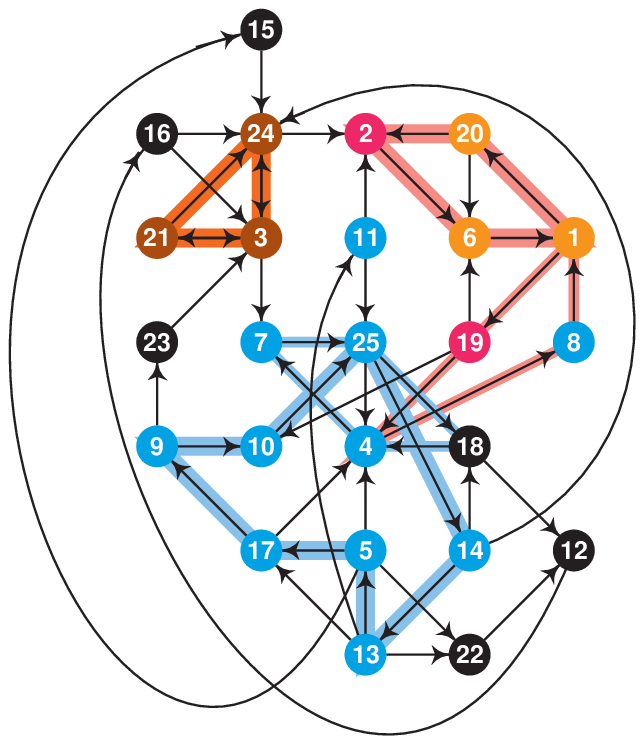,width=2.5in}}
$~~$
\subfigure[\label{units}Three cyclic units of \lscc with connections.]
{\epsfig{file=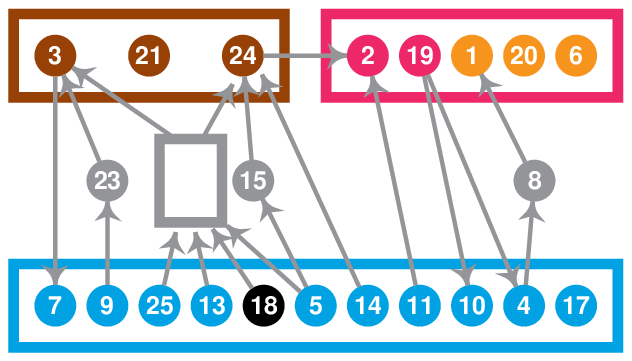,width=2in}}
}
\caption{
At least three feedback vertices are needed because there exists three
vertex-disjoint cycles --- indicated by wide color strips.  If a single
vertex selection on an indicated cycle suffices for the feedback vertex set
then it must intersect every cycle that is vertex disjoint with the other
indicated cycles; cycles indicated with thin color strips show that
such selections are unique.
}
\end{figure}

Another property of \lscc 
is that it has a small and unique minimum {\em feedback vertex set},
a set of nodes whose removal destroys all cycles.

The fact that there exists a unique minimum feedback vertex set
with three nodes (vertices) can be clearly seen in \FI{fvs}.  Let us
call this set $F=\{1,3,25\}$.

We can use $F$ to distinguish three natural cyclic units within \lscc,
$S_b$ for each $b\in F$.
We can think that $b$ is the ``boss'' of $S_b$.  We define $S_b$ as
the union of all simple cycles that go through $b$ but not
through $F-\{b\}$.  Only one node can have two bosses: $\{4\}=S_1\cap S_{25}$.
Because there is only one path from 1 to 4 and three disjoint paths
from 25 to 4, we remove 4 from $S_1$ to make our units disjoint.
The three sets coincide well with
functional categories: $S_3=\{3,21,24\}$
are the nodes on cycles of $\Lsc{sr}$,
$S_1$ are the nodes on cycles of $\Lsc{sp}$, and $S_{25}$
are the nodes on cycles of $\Lsc{cc}$ minus $S_1$ (observe that
$S_1$ is contained in $\Lsc{cc}$).
(Actually,
$S_{25}$ has 11 nodes and it has one node that is not in $\Lsc{cc}$,
18, and one node of the cell cycle network is missed, 8.)

Thus the cyclic subnework has three cyclic parts, plus two acyclic parts:
5 nodes on paths from $S_{25}$ to $S_3$, and 1 node on a path from
$S_{25}$ to $S_1$.
We show this schematically in \FI{units}.
These units are related to ``large network structures'' that were observed, but
not described, by Lee \EA{L02}.

\subsection{Statistic profile of the TFs from the \lscc}

We tested 1000 random networks generated in three ways:
\begin{enumerate}
\item
with the same in- and out-degrees as in the actual network;
\item
the same, but with the number of bi-fans increased to the
actual using random swaps of edge ends, and accepting them
when they increase the number of bi-fans;
\item
similar to the last one, but increasing the number of
feed-forward loops.
\end{enumerate}

These three methods yielded similar results.  We will be
reporting these results in the format $a$ ($b,c$) where
$a,b,c$ will be the averages obtained using method 1, 2 and 3 respectively.

\subsubsection{Average out-degree in the \lscc.}
On the average, a transcription factor has 50 targets, and the average
for the transcription factors of the \lscc is 128.
This agrees with the average 121 (120, 123) for the random model.
It can be explained by the nodes of large out-degree having
much larger chances of joining the cycles.
In the actual network, of the top 20 most active
TFs, the \lscc has 12.

\subsubsection{Average co-regulation in the \lscc and overall.}

We define the co-regulation of two
TFs as the number of shared target genes.  If two TFs have $t_1$ and
$t_2$ targets, while the total number of targets is $n$, on the
average they share $E=t_1t_2/n$ targets.  If they actually share
$A$ targets, their relative co-regulation is $A/E$.
Over all pairs of TFs, the average relative co-regulation is
2.93, and for the pairs in the \lscc the average
is 2.0.  The explanation is that the {\em relative} co-regulation
tends to be high for the TFs with small number of targets (when
the expected co-regulation is very small).  If we increase the
number of bi-fans by random re-wiring, the average co-regulation
increases modestly, because the gains occur mostly for the  TFs
with a large number of targets.  Therefore the ``generating force''
of bi-fans is not the random re-wiring.  Mutation by duplication (see
Teichmann and Babu \cite{T04})
can explain this pattern --- a duplicated pair has large co-regulation
even if it has small out-degree.

In the random networks, the relative co-regulation is smaller.  When
we boost the number of bi-fans, this should increase co-regulation.
However, it is much easier to increase the number of bi-fans at
random for a node with large out-degree, and as a result, this process
actually decreases the average relative co-regulation (to 0.87 from 1.27).
This test does show that the distribution of bi-fans, relative to the
out-degree of the participating nodes, is very different than in a
random network with the same number of bi-fans.

\begin{table}[t]
\epsfig{file=table_hi.eps,width=2.7in}~$~~$~\parbox{2in}{
\vspace{-48ex}
The paths are computed in the graph of \scc's, in particular,
we view \lscc as a single node.
The entry in column $i$ and row $j$ shows the number of nodes with these
properties: on the longest path through node $u$ has $i+j$ edges and
the longest path from $u$ {\bf to} another node (a TT) has $i$ edges
(consequently, the longest path {\bf from} another node to $u$ has $j$ edges.) 
Note that the only way a node may be on a path of length 3
is when it has an edge from the node that corresponds to \lscc.
}
\caption{\label{hier}
Classifying TFs and TTs by their positions on the longest paths.
}
\end{table}

\begin{table}[t]
\begin{tabbing}
56789
\=01234567890123\=0123456789\=0123456789\=0123456789\=0123456789\=0123456789\=\kill
\> \> Top \> Cycle\>\hspace{-1ex}Complex\> Simple\> Exception \\
\> Top (9)       \> 168 \> \ 373 \> \ 373 \> \ 121 \> \ \ 12 \\
\> Cycle (25)    \>     \>  1132 \> \ 696 \> \ 249 \> \ \ 21 \\
\> Complex (65) \>     \>       \> \ 638 \> \ 259 \> \ \ 31 \\
\> Simple (38)  \>     \>       \>       \> \ 169 \> \ \ 57 \\
\> Exception (2) \>     \>       \>       \>       \> \ \ 14
\end{tabbing}
\caption{\label{corhier}Co-regulation of various hierarchy classes}

\end{table}

\subsubsection{Position of \lscc in the hierarchy.}
Only 9 TFs are ``upstream'' from the \lscc in the sense
that there are paths from these TFs to the \lscc; of these 9 paths
8 are single edges and one consists of two edges.
If we consider
that path to be exception, collectively the cyclic component has
unambiguous hierarchical position 2nd from the top.
In a random network, on the average we have 17 (16.8, 14.8)
``upstream'' TFs.  In this
sense, the cyclic component is higher in the hierarchy than the
average in the random model.

In the random model we can see that most of the long paths go through
the large cyclic component, in the actual network this is even more so.
Every TF (with two exceptions) which is on a path of length 3
or more either belongs to the cyclic component, or it can reach the cyclic
component, or it can be reached from it.  This means that 38 TFs are on very
short paths only and form a rather separate part of the transcription network,
while 104 TFs belong to paths of length 3 or more.  The length
of the longest path measured when we collapse the cyclic component to
single nodes is 13 in the actual network, and on the average 9.4 (9.2, 9.4).
\footnote{The maximum length of a simple path is perhaps a better measure,
but it requires a much more complex program to compute it.  It is
closely related to the feedback vertex set problem.}

Yu and Gerstein \cite{Y06}
propose a partition of networks according to the length of shortest
paths to those TFs that have only TTs as their targets.  This definition would not
work with the length of the shortest paths to TTs: this length is 1 for all TFs
but ten, and for that ten, it is 2, so the hierarchy would be trivial.

Because \lscc has such a special and statistically
significant position in the network, we propose to partition TFs by their
relation to \lscc, as it is indicated in \TA{hier}.

\TA{corhier} considers five classes of TFs from \TA{hier},
the fifth class consisting
of two TFs that do not fit into our schema.  For each pair of classes
we give the number of TTs that are co-regulated by TFs from those two
classes (positions like Top-Top give the number of TTs regulated by
that class alone, the size of each class is given in parenthesis).

We performed our study using the data of
Luscombe \EA{L04}
because we wanted to compare the cycles with physiological subnetworks
described in their paper.  Nevertheless, we compared our definition
of a hierarchy with that of
Yu and Gerstein \cite{Y06},
who performed their investigation in a larger transcription network.

When we apply our program to the latter network,
the proportions between the class sizes remain similar:
Top (20), Cycle (63), Complex (114), Simple (84) and Exception (5).

We performed two tests applied by Yu and Gerstein to their classes.
When we checked the percentage of essential genes
in our classes, we got 15\% in Top and Cycle, 13\% in Complex and 12\% in
Simple, a more uniform distribution than among classes of Yu and Gerstein.
A more striking difference exists when we check the percentage of
cancer related genes: 25\% in Top, 16\% in Cycle, 12\% in Simple and
below 3\% in Complex.

The division we propose is closely related to the notion proposed by
Yu and Gerstein: a division of transcription control mechanisms into
{\em reflex} processes and {\em cogitation} processes.  Simple clearly
corresponds to reflex processes.  In a cogitation process, one that
involves a long path of interactions, we can partition the process into
beginning, middle and the ending part.  As the various paths have very
different lengths, identifying \lscc as the middle is both ``objective'' and
independent
from the path length, and in the same time quite arbitrary.  However, we
show in the next subsection that \lscc has a ``switchboard'' property even
in the physiological conditions in which paths do not form cycles, and
we just have seen that
the percentage of cancer related genes sharply drops as we move from the
middle to the final part of the long paths.

\subsection{Topological changes inside \lscc}
\label{topochanges}

\begin{figure}[t]
\centerline{\epsfig{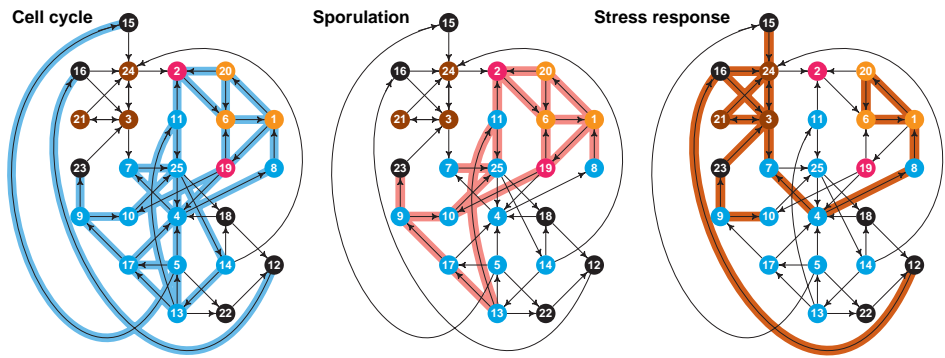}}
\caption{\label{changeso}
Parts of \lscc that are active during endogenous condition
(or, conditions with larger number of active cycles). 
}
\end{figure}

\begin{figure}[t]
\centerline{\epsfig{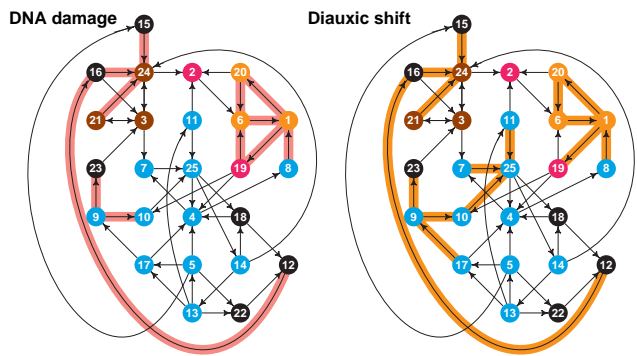}}
\caption{\label{changest}
Parts of \lscc that are active during exogenous
condition (or, conditions with the fewest active cycles).
}
\end{figure}

In \FI{changeso} and \FI{changest}
we can see the interactions of \lscc that
are active in various physiological conditions.  We can observe
large difference between the subnetworks, both in the composition
and in topological characteristics like average path length.

Because so many paths of TF go through \lscc, the differences between
average path lengths that were observed for different subnetworks 
by
Luscombe \EA{L04}
are largely caused by the different presence of these networks
in the \lscc.  In \TA{topol} we use {\sc PercentPath} to denote the percentage of the shortest
paths from transcription factors to the terminal targets
that either originate or go through \lscc, and {\sc PercentLength}
to denote the similar percentage for the sum of lengths of shortest paths.

\TA{topol} shows
that even in DNA damage and diauxic shift subnetworks the majority
of shortest paths between TFs and TTs goes throgh \lscc; we may say that
\lscc has a role of a {\em switchboard}.

\begin{table}[t]
\begin{tabbing}
0123456789
\=12345678901234567890\=1234567\=1234567\=1234567\=1234567\=1234567\=\kill
\> subnetwork\> \ {\bf cc}\> \ {\bf sp}\> \ {\bf sr}\> \ {\bf ds}\> \ {\bf dd}\\
\> average path length \> 4.64 \> 3.55 \> 2.31 \> 2.10 \> 1.94 \\
\> {\sc PercentPath} \> 87.1 \> 69.4 \> 72.1 \> 57.8 \> 54.6 \\
\> {\sc PercentLength} \> 94.2 \> 78.0 \> 81.6 \> 64.4 \> 59.0 \\ \end{tabbing}
\caption{\label{topol}Importance of \lscc in the paths of different subnetworks}
\end{table}

\subsubsection{Position of \lscc in the hierarchy.}
Only 9 TFs are ``upstream'' from the \lscc in the sense
that there are paths from these TFs to the \lscc; of these 9 paths
8 are single edges and one consists of two edges.

\vspace{-1ex}
\section{Conclusions}

\vspace{-1ex}
We inspected graph-theoretic properties of the cycles in
the transcription network of \Sac.  While in general cycles
are ``avoided'' by the network, interactions common to all
phases of the cell cycle form a big exception, and interactions
specific to the stress response form a smaller exception.

In spite of their modest number (they involve 25 of 142 transcription factors
that were included in the data set), the transcription factors that are
included in cycles have a large topological impact: most of the shortest paths
between transcription factors and terminal targets go through them.

One should compile many kinds of data to establish the exact role
of the cycles of transcription interactions in controlling life processes. 
In particular, cell cycle, which is closely related to cancer, possesses
a long cycle that can be easily interrupted at many different points, and
the process itself can be interrupted by a number of different conditions
(like DNA damage).

We have shown that \lscc is a key part of the regulatory network and that
it can be divided into functional subunits.  Further work will yield
fuller and clearer picture of these subunits and their interactions
under various conditions.
\section{Methods}

\vspace{-1ex}
\subsection{Data}
We used supplementary materials for
\cite{L04}
(at {\tt http://sandy.topnet.gersteinlab. org/index2.html}); we also used
supplementary materials of \cite{Y06} and the list of yeast homologs of
human cancer genes personally communicated by Haiyuan Yu.
\vspace{-2ex}
\subsection{Graph-theoretic definitions}
A {\em graph} of a network consists of {\em nodes} (which correspond
to TFs, transcription factors and TTs, terminal targets) and directed
edges/interactions.

A {\em path} in a graph is a sequence of nodes $(u_0,\ldots,u_{k-1})$ such that
each consecutive pair $(u_{i-1},u_i)$ is an edge.  If additionally there
exists an edge $(u_{k-1},u_0)$ we say that this is a {\em cycle}.

A single node $(u)$ forms a {\em degenerate} cycle.

Nodes in a graph are partitioned into {\em strongly connected components},
or \scc's.  A node $u$ is contained in $\scc(u)$ which 
is the union of the node sets of all cycles that contain $u$.

\scc's with one node are called trivial.

For graph $G$ we define graph $G_{\scc}$, the graph of \scc's of $G$.
Nodes of $G_{\scc}$ are scc's of $G$, and edges are pairs of the form
$(\scc(u),\scc(v))$ such that $(u,v)$ is an edge of $G$.

$G_{\scc}$ cannot have cycles of its own, and therefore it is easy to
compute longest paths in that graphs (the algorithm is considered folklore).
The paths lengths in that graph are used in \TA{hier}.

We use \lscc to denote the largest strongly connected component in a graph.
We apply this definition when the majority of elements of non-trivial
\scc's belongs to one of them, so there is no ambiguity as to which one is
``the largest''.
\vspace{-2ex}
\subsection{Algorithms}
To compute non-trivial scc's we first
obtained a ``dictionary'' protein code $\leftrightarrow$ number followed
by pairs of numbers representing the edges.
We computed \scc's and the graph of \scc's using the
algorithm of Tarjan \cite{Ta72}.  His method is usually described
in textbooks of algorithms for {\em biconnected components} (the difference
between two algorithms is contained in one line of code).
Shortest paths used in subsection \ref{topochanges}
were computed using breadth first search.

We implemented this algorithm in two programming languages: in awk, which
makes it very easy to compare the result with various text files, and in
C which allows to perfom very quickly thousands of statistical tests.
\vspace{-2ex}
\subsection{Defining motifs, generating random graphs}
We define a feed-forward loop (ffl for short) as a triple of nodes
$\{u_0,u_1,u_2\}$ such that there exists three edges: two form a path
$(u_0,u_1,u_2)$ while the third forms a shortcut, $(u_0,u_2)$.
A bi-fan is a quadruple of nodes $(u_0,u_1,v_0,v_1)$ such that all
of the 4 possible edges of the form $(u_i,v_j)$ exist.

When we count ffl's and bi-fans we remove the self-loops (edges of the
form $(u,u)$) from the graph.  Moreover, every triple/quadruple is counted
separately, even when they share nodes.

To count ffl's and bi-fans we made a table Overlap that for a pair of
TFs stored the number of common targets.  For every positive entry
$k = Overlap(a,b)$ we add $k(k-1)/2$ to the count of bi-fans, and if
there is an edge from $a$ to $b$, we add $Overlap(a,b)$ to the count of ffl's.

The method of generating random graphs was adapted from
Milo \EA{M03}.
For a node $u$ with in-degree $a$ and out-degree $b$ we conceptually
make $a$ ``in-stubs'' and $b$ ``out-stubs'', in actuality, we have
an array in which $b$ positions are reserved for the adjacency list
of $u$, and $u$ is in $a$ locations of that array.  Then we pick a random
permutation of the array content.  Subsequently, we sort the adjacency lists.
Finally, we scan the adjacency lists and we correct ``errors'' which are
of two types: a list of some node $v$ contains $u$ for the second time, or
it contains $v$ itself.  We try to exchange the offending array position
with a randomly chosen other position until the exchange does not introduce
an error of its own.

Similar approach is used to ``boost'' the number of bi-fans or ffl's: a random
swap of two array positions is accepted if it does not introduce an
error and it increases the respective count (ffl's to 997 or bi-fans to
61,034).

\vspace{-2ex}
\section*{Acknowledgements}
\vspace{-2ex}
The author is grateful to Arthur Lesk for posing the problem and
reviewing the manuscript.  Discussions
with Arthur Lesk, L. Aravind and Piotr Berman were very helpful.
Haiyuan Yu supplied one of the data sets.  
\vspace{-2ex}
 
\bibliographystyle{plain}
\bibliography{jeong_cycle}

\begin{thebibliography}{10}

\bibitem{B99}
A.~L. Barabasi and R.~Albert.
\newblock Emergence of scaling in random networks.
\newblock {\em Science}, 286:509--512, 1999.

\bibitem{La07}
M.~Cosentino~Lagomarsino, P.~Jona, B.~Bassetti, and H.~Isambert.
\newblock Hierarchy and feedback in the evolution of the escherichia coli
  transcription network.
\newblock {\em Proc. Natl. Acad. Sci.}, 104:5516--5520, 2007.

\bibitem{H02}
L.H. Hartwell.
\newblock Yeast and cancer.
\newblock {\em Bioscience Reports}, 22:372--394, 2002.

\bibitem{L04}
N.M. Luscombe, M.M. Babu, H.~Yu, M.~Snyder, S.A. Teichmann, and M.~Gerstein.
\newblock Genomic analysis of regulatory network dynamics reveals large
  topological changes.
\newblock {\em Nature}, 431:308--312, 2004.

\bibitem{M93}
W.H. Mager and P.M. Ferreira.
\newblock Stress response of yeast.
\newblock {\em Biochem J.}, 290:1--13, 1993.

\bibitem{M03}
R.~Milo, N.~Kashtan, S.~Itzkovitz, M.E.J. Newman, and U.~Alon.
\newblock On the uniform generation of random graphs with prescribed degree
  sequences, 2003.

\bibitem{N01}
M.E.J. Newman, S.H. Strogatz, and D.J. Watts.
\newblock Random graphs with arbitrary degree distributions and their
  applications.
\newblock {\em Physical Review E}, 64:026118, 2001.

\bibitem{A02}
S.S. Shen-Orr, R.~Milo, S.~Mangan, and U.~Alon.
\newblock Network motifs in the transcriptional regulatory network of
  escherichia coli.
\newblock {\em Nature Genetics}, 31:60--63, 2002.

\bibitem{Ta72}
R.E. Tarjan.
\newblock Depth first search and linear time algorithms.
\newblock {\em SIAM J. of Computing}, 1:146--160, 1972.

\bibitem{T04}
S.A. Teichmann and M.M. Babu.
\newblock Gene regulatory network growth by duplication.
\newblock {\em Nature Genetics}, 36:492--496, 2004.

\bibitem{L02}
I.~L. Tong, N.~J. Rinaldi, F.~Robert, D.~T. Odom, Z.~Bar-Joseph, G.~K. Gerber,
  N.~M. Hannett, C.~T. Harbison, C.~M. Thompson, I.~Simon, J.~Zeitlinger, E.~G.
  Jennings, H.~L. Murray, D.~B. Gordon, B.~Ren, J.~J. Wyrick, J.-B. Tagne,
  Th.~L. Volkert, E.~Fraenkel, D.~K. Gifford, and R.~A. Young.
\newblock Transcriptional regulatory networks in saccharomyces cerevisiae.
\newblock {\em Science}, 298:799--804, 2002.

\bibitem{Y06}
H.~Yu and M.~Gerstein.
\newblock Genomic analysis of the hierarchical structure of regulatory
  networks.
\newblock {\em Proc. Natl. Acad. Sci.}, 103:14724--31, 2006.

\end{thebibliography}

\end{document}